\documentclass[a4paper,english,runningheads]{llncs}
\usepackage[T1]{fontenc}
\usepackage[latin9]{inputenc}
\usepackage{float}
\usepackage{textcomp}
\usepackage{amsmath}
\usepackage{amssymb}
\usepackage{graphicx}

\makeatletter

\providecommand{\tabularnewline}{\\}

\usepackage[all]{xy}

\usepackage{cite}
\usepackage{babel}

\newcommand{\sodd}{\mathsf{s}_{1}}
\newcommand{\ICC}{\text{ICC}}

\newcommand{\random}[1]{#1}      
\renewcommand{\random}[1]{\mathit{\uppercase{#1}}}

\newcommand{\sets}[1]{#1}        
\renewcommand{\sets}[1]{\mathrm{\uppercase{#1}}}

\newcommand{\unrelated}[1]{#1}   
\renewcommand{\unrelated}[1]{\mathfrak{\uppercase{#1}}}

\newcommand{\eqDist}{\buildrel d \over =}

\makeatother

\usepackage{babel}
\begin{document}

\title{Exploration of Contextuality in a Psychophysical Double-Detection
Experiment}

\titlerunning{Contextuality in Double-Detection}

\author{Víctor H. Cervantes\textsuperscript{1}, Ehtibar N. Dzhafarov\textsuperscript{2}}

\authorrunning{V. H. Cervantes, E. N. Dzhafarov}

\institute{\textsuperscript{1}Purdue University\\
 cervantv@purdue.edu\\
 $\,$\\
 \textsuperscript{2}Purdue University\\
 ehtibar@purdue.edu}

\titlerunning{Contextuality in Double-Detection}

\tocauthor{V. H. Cervantes, E. N. Dzhafarov}
\maketitle
\begin{abstract}
The Contextuality-by-Default (CbD) theory allows one to separate contextuality
from context-dependent errors and violations of selective influences
(aka ``no-signaling'' or ``no-disturbance'' principles). This
makes the theory especially applicable to behavioral systems, where
violations of selective influences are ubiquitous. For cyclic systems
with binary random variables, CbD provides necessary and sufficient
conditions for noncontextuality, and these conditions are known to
be breached in certain quantum systems. We apply the theory of cyclic
systems to a psychophysical double-detection experiment, in which
observers were asked to determine presence or absence of a signal
property in each of two simultaneously presented stimuli. The results,
as in all other behavioral and social systems previous analyzed, indicate
lack of contextuality. The role of context in double-detection is
confined to lack of selectiveness: the distribution of responses to
one of the stimuli is influenced by the state of the other stimulus.

\end{abstract}

The Contextuality-by-Default (CbD) theory \cite{Dzhafarov.2016.Contextuality-by-Default:,Dzhafarov.2016.contextuality}
describes systems of measurements with respect to the conditions under
which they are recorded and determines the tenability of a non-contextual
description of the system. In this paper, we study the double-detection
paradigm suggested in Refs.~\cite{Dzhafarov.2012.Quantuma} and~\cite{Dzhafarov.2013.Probability}.
In this paradigm, two stimuli are presented to an observer simultaneously
(left-right), each on one of several possible levels. The observer
is asked to state (Yes/No), for each of the two observation areas,
whether it contains a particular target property (signal). The signal
is objectively present in a subset of levels of a stimulus. When such
experimental situation includes only two levels for each stimulus
(e.g., present/absent), the system of measurements is formally equivalent
to that of the Einstein\textendash Podolski\textendash Rosen/Bohm
(EPR/B) paradigm (see e.g, Ref.~\cite{Dzhafarov.2012.Quantuma}).

\section{Contextuality in CbD}

\label{sec:intro}

We briefly recapitulate the concepts of the CbD, to make this paper
self-sufficient. For detailed discussions see Ref.~\cite{Dzhafarov.2016.Contextuality-by-Default:}
and Ref.~\cite{Dzhafarov.2016.contextuality}; the proofs may be
found in Refs.~\cite{Dzhafarov.2015.Contextualitya,Kujala.2015.Proofa,Kujala.2016.Probabilistic}.

\begin{definition}\textbf{\emph{(System of measurements)}}\label{def:system}
A \emph{system of measurements} is a matrix $\unrelated{R}_{n\times m}$,
in which columns correspond to the properties $\left\{ q_{1},\ldots,q_{n}\right\} $
and rows to the contexts $\left\{ c_{1},\ldots,c_{m}\right\} $. A
cell $\left(i,j\right)$ contains the random variable $R_{i}^{j}$
if $q_{i}$ is measured in context $c_{j}$, and the cell is left
empty otherwise. \end{definition}

When adopting the CbD framework, the first goal is to produce a matrix
$\unrelated{R}$ that formally represents the experiment and its results. 

\begin{definition}\textbf{\emph{(Connections and bunches)}}\label{def:Connection}
The random variables in any column of a system of measurements form
a \emph{connection} for the corresponding property; denote the connection
for property $q_{i}$ by $\unrelated{R}_{i}$. Those in any row form
a \emph{bunch} representing the corresponding context; denote the
bunch for context $c_{j}$ by $\random{R}^{j}$.

\end{definition}

Note that elements of a connection are necessarily (``by default'')
pairwise distinct and pairwise stochastically unrelated, i.e., no
$R_{i}^{j}$ and $R_{i}^{k}$ with $k\not=j$ have a joint distributions.
Consequently, the system $\unrelated{R}$ does not have a joint probability
distribution including all of its elements. See Refs.~\cite{Dzhafarov.InPress.Stochastic,Dzhafarov.2016.contextuality}. 

\begin{definition}\textbf{\emph{(Coupling)}}\label{def:coupling}
Let $\random{X}_{i}$, with $i\in\sets{I}$, an index set, be a random
variable on a probability space $(\sets{X}_{i},\Sigma_{i},P_{i})$.
Let $\left\lbrace \random{Y}_{i}:i\in\sets{I}\right\rbrace $ be a
collection of jointly distributed random variables (i.e., a random
variable in its own right) on a probability space $(\sets{Y},\Omega,p)$.
The random variable $\left\lbrace \random{Y}_{i}:i\in\sets{I}\right\rbrace $
is called a \emph{coupling} of the collection $\left\{ \random{X}_{i}:i\in\sets{I}\right\} $
if for all $i\in\sets{I}$, $\random{Y}_{i}\eqDist\random{X}_{i}$,
where $\eqDist$ denotes identity in distribution. 

\end{definition}

\begin{definition}\textbf{\emph{(Maximal coupling)}}\label{def:maxCoupling}
Let $\random{Y}=\left(\random{Y}_{i}:i\in\sets{I}\right)$ be a coupling
of a collection $\left\{ \random{X}_{i}:i\in\sets{I}\right\} $. And
let $\sets{M}$ be the event where $\left\lbrace \random{Y}_{i}=\random{Y}_{j}\text{ for all }i,j\in\sets{I}\right\rbrace $.
If $\Pr(\sets{M})$ is the largest possible among all couplings of
$\left\{ \random{X}_{i}:i\in\sets{I}\right\} $, then $\random{Y}$
is a \emph{maximal coupling} of $\left\{ \random{X}_{i}:i\in\sets{I}\right\} $.

\end{definition}

\begin{definition}\textbf{\emph{(Contextual system)}}\label{def:contextualSystem}
Let $\unrelated{R}$ be a system of measurements. Let $\random{S}$
be a coupling of $\unrelated{R}$ such that for each $c_{j}\in\left\{ c_{1},\ldots,c_{m}\right\} $,
$\random{S}^{j}$ is a coupling of $\random{R}^{j}$ contained in
$\random{S}$. The system $\unrelated{R}$ is said to be \emph{non-contextual}
if it has a coupling $\random{S}$ such that for all $q_{i}\in\left\{ q_{1},\ldots,q_{n}\right\} $,
the coupling $\random{S}_{i}$ is a maximal coupling. 

\end{definition}

\begin{definition}\textbf{\emph{(Cyclic system with binary variables)}}\label{def:cyclicSystem}
Let $\unrelated{R}$ be a system of measurements such that (a) each
context contains two properties; (b) each property is measured in
two different contexts; (c) no two contexts share more than one property;
and (d) each measurement is a binary random variable, with values
$\pm1$. Then the system $\unrelated{R}$ is a \emph{cyclic system
with binary variables} and in the following will be simply called
\emph{a cyclic system}. \end{definition}

\begin{remark}Note that a cyclic system is composed of the same number
$n$ of connections and of bunches, and it contains $2n$ random variables.
We shall say that a cyclic system has rank $n$ or is of rank $n$
to explicitly refer to this number.

\end{remark}

\begin{definition}\textbf{\emph{(Consistent connections)}}\label{def:consistentConnection}
Let $\unrelated{R}_{i}$ be a connection in a system $\unrelated{R}$.
It is said that $\unrelated{R}_{i}$ is a \emph{consistent connection}
if for all $c_{j},c_{k}\in\left\{ c_{1},\ldots,c_{m}\right\} $ such
that $\random{R}_{i}^{j}$ and $\random{R}_{i}^{k}$ are defined (i.e.,
both cells $(i,j)$ and $(i,k)$ of $\unrelated{R}$ are not empty),
$\random{R}_{i}^{j}\eqDist\random{R}_{i}^{k}$. \end{definition}

\begin{definition}\textbf{\emph{(Consistently connected system)}}\label{def:consistentSystem}
A system of measurements $\unrelated{R}$ is said to be \emph{consistently
connected} if for all $q_{i}\in\left\{ q_{1},\ldots,q_{n}\right\} $,
the connection $\unrelated{R}_{i}$ is a consistent connection. For
a cyclic system, define 
\[
\ICC=\sum_{i=1}^{n}\left|\left\langle \random{R}_{i}^{j}\right\rangle -\big\langle\random{R}_{i}^{k}\big\rangle\right|.
\]
\ICC{} provides a measure of how inconsistent the connections are
in the system. 

\end{definition}

\begin{definition}\textbf{\emph{(Contextuality in cyclic systems)}}\label{def:cyclicMeasure}
Let $\unrelated{R}$ be a cyclic system with $n$ binary variables.
Let 
\[
\sodd(x_{1},x_{2},\ldots,x_{n})=\max\left\lbrace \sum_{k=1}^{n}a_{k}x_{k}:a_{k}=\pm{1}\text{ and }\prod_{k=1}^{n}a_{k}=-1\right\rbrace .
\]
Let 
\[
\Lambda{C}=\sodd\left(\left\lbrace \left\langle \random{R}_{i}^{j}\random{R}_{i'}^{j}\right\rangle :q_{i},q_{i'}\text{ measured in }c_{j},\text{ and }c_{j}\in\left\{ c_{1},\ldots,c_{m}\right\} \right\rbrace \right)
\]
Let $\Delta{C}=\Lambda{C}-\ICC-(n-2)$. The quantity $\Delta{C}$
is a \emph{measure of contextuality} for cyclic systems. 

\end{definition}

\begin{theorem} \textbf{\emph{(Cyclic system contextuality criterion,
\cite{Kujala.2015.Proofa})}}\label{the:criterionCyclic} A cyclic
system $\unrelated{R}$ is contextual if and only if $\Delta{C}>0$.
\end{theorem}

\begin{remark} $\Delta{C}$ for a consistently connected cyclic system
with $n=4$ reduces to the Bell/CHSH inequalities \cite{Clauser.1969.Proposed,Dzhafarov.2016.contextuality}.
\end{remark}

\section{Contextuality in Behavioral and Social Data}

\label{sec:behavior}

In Ref.~\cite{Dzhafarov.2015.there} many empirical studies of behavioral
and social systems were reviewed. Most of those systems come from
social data; that is, an observation for each measurement was the
result of posing a question to a person, and the set of observations
comes from questioning groups of people. For all the studies considered
there, the CbD analyses showed that the systems, treated as cyclic
systems ranging from rank 2 to 4, were non-contextual. Only one of
the studies reviewed in Ref.~\cite{Dzhafarov.2015.there} dealt with
responses from a single person to multiple replications of stimuli.

Now, a key modeling problem in cognitive psychology has been determining
whether a set of inputs selectively influences a set of response variables
(Refs.~\cite{Sternberg.1969.discovery,Townsend.1984.Uncovering,Dzhafarov.2003.Selective,Zhang.2015.Noncontextuality}).
The formal theory of selective influences has been developed for the
case of consistent connectedness, which has been treated as a necessary
condition of selective influences; it follows from this formalism
that selectiveness of influences in a consistently connected system
is negated precisely in the case where it is contextual \cite{Dzhafarov.2012.Quantuma}.

However, in most, if not all, behavioral systems, some form of influence
upon a given random output is expected from most, if not all, of the
system's inputs (Ref.~\cite{Townsend.1984.Uncovering}). This means
that in the behavioral domain inconsistently connected systems are
ubiquitous. While the presence of inconsistent connections rules out
the possibility of selective influences, it does not imply that the
full behavior of the system is accounted for by the direct action
of inputs upon the outputs; an inconsistently connected behavioral
system may still be contextual in the sense of CbD.

The double detection paradigm suggested in \cite{Dzhafarov.2012.Quantuma}
and \cite{Dzhafarov.2013.Probability} provides a framework where
both (in)consistent connectedness and contextuality can be studied
in a manner very similar to how they are studied in quantum-mechanical
systems (or could be studied, because consistent connectedness in
quantum physics is often assumed rather than documented). 

\section{Method}

\label{sec:method}

\subsection{Participants}

\label{sec:part} Three volunteers, two females and one male, graduate
students at Purdue University, served as participants for the experiment,
including the first author of this paper. They were recruited and
compensated in accordance to Purdue University's IRB protocol $\#1202011876$,
for the research study ``Selective Probabilistic Causality As Interdisciplinary
Methodology'' under which this experiment was conducted. All participants
reported normal or corrected to normal vision and were aged around
$30$. They are identified as $P1-P3$ in the text and their experience
with psychophysical experiments ranged from none to more than three
previous participations.

\subsection{Apparatus}

\label{sec:apparatus} The experiment was run using a personal computer
with an Intel\textsuperscript{{\scriptsize{}\textregistered{}}} Core{\scriptsize{}\texttrademark{}}
processor running Windows XP, a $24$-in.\ monitor with a resolution
of $1920\times1200$ pixels (px), and a standard US $104$-key keyboard.
A chin-rest with forehead support was used so that the distance between
subject and monitor was kept at $90$ cm; this made each pixel on
the screen to occupy about $62$ sec arc of the subjects' visual field.

\subsection{Stimuli}

\label{sec:stim} 
\begin{figure}[tbh]
\centering{}\includegraphics[width=0.78\textwidth]{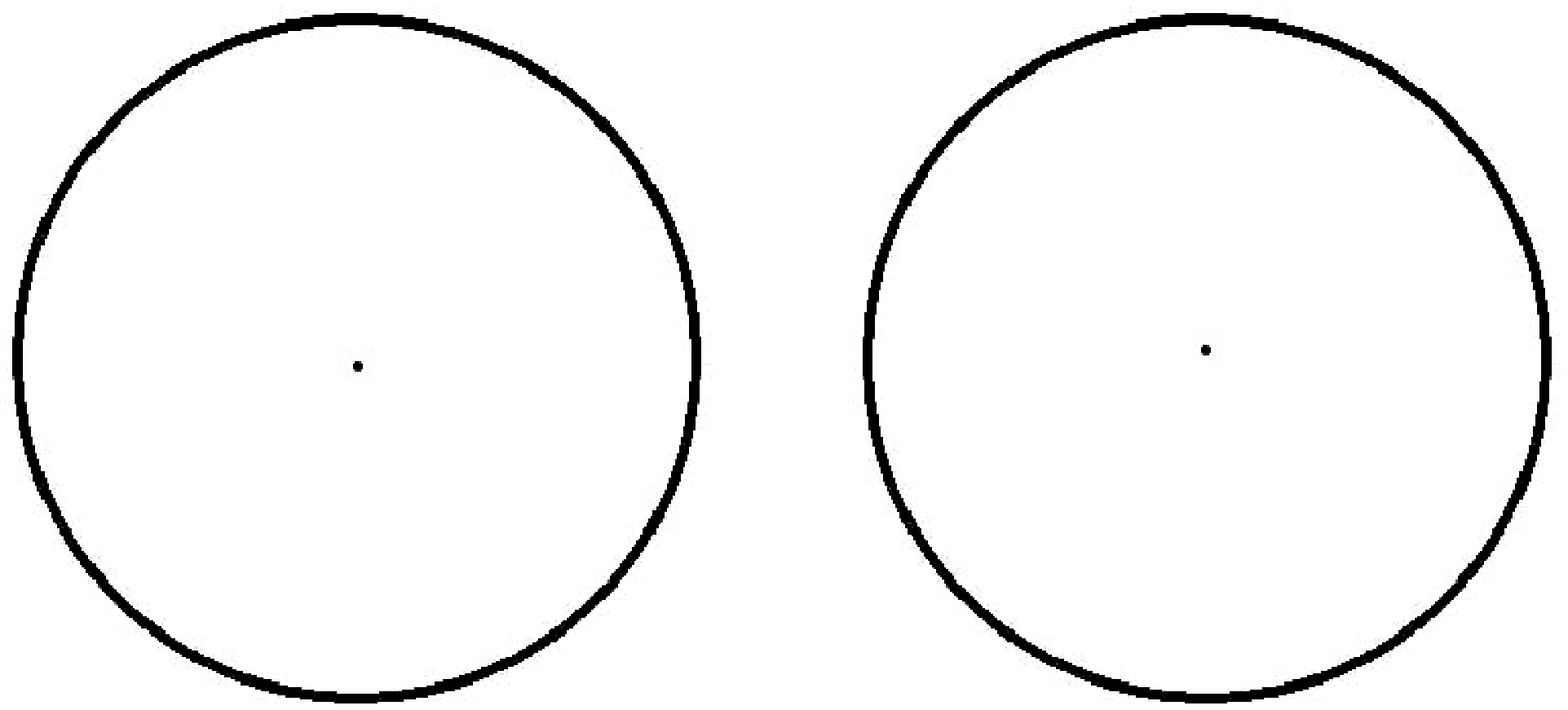}
\caption{Stimulus example}
\label{fig:stimulus} 
\end{figure}

The stimuli were similar to those from Refs.~\cite{Allik.2014.Detection}
and \cite{Dzhafarov.2010.Matching}. They consisted of two circles
drawn in solid grey (RGB $100$, $100$, $100$) on a black background
in a computer screen, with a dot drawn at or near their center. The
circles radius was $135$ px with their centers $320$ px apart; the
dots and circumference lines were $4$ px wide. The offset of each
dot with respect to the center of each circle, when they were not
presented at the center, was $4$ px. An example of the stimuli (in
reversed contrast) is shown in Figure~\ref{fig:stimulus}. 

\subsection{Procedure}

\label{sec:proc} Each participant performed nine experimental sessions.
At the beginning of each experimental session, the chin-rest and chair
heights were adjusted so that the subject could sit and use the keyboard
comfortably. The time available for each session was $30$ minutes,
during which the participants responded in $560$ (non-practice) trials
(except for participant $P3$ in the sixth session, who only responded
in $557$ trials) preceded by up to 30 practice trials. The number
of practice trials was set to $30$ during the first two sessions
and reduced to $15$ during subsequent sessions. After each practice
trial, the subject received feedback about whether their response
for each circle was correct or not. The responses to practice trials
were excluded from the analyses. Additionally, depending on their
previous experience in psychophysical experiments the participants
had up to three training sessions, also excluded from subsequent analyses.

Instructions for the experiment were presented to each participant
verbally and written in the screen. In each trial the participant
was required to judge for each circle whether the dot presented was
displaced from the center or not. The stimuli were displayed until
the subject produced their response. The responses were given by pressing
and holding together two keys, one for each circle. Then, the dots
in each circle were removed and a ``Press the space bar to continue''
message was flashed on top of the screen. After pressing the space
bar, the message was removed and the next stimuli pair were presented
after $400$ ms. (Reaction times were measured from the onset of stimulus
display until a valid response was recorded, but they were not used
in the data analysis.)

\subsection{Experimental Conditions}

\label{sec:expCond} In each of two circles the dot presented could
be located either at its center, or 4 px above, or else 4 px under
the center. These locations produce a total of nine experimental conditions.

During each session, excepting the practice trials, the dot was presented
at the center in a half of the trials; above the center in a quarter
of them; and below the center in the remaining quarter, for each of
the circles. Table~\ref{tab:allocation} presents the proportions
of allocations of trials to each of the 9 conditions.

\begin{table}[tph]
\caption{Probabilities with which a trial was allocated to one of the 9 experimental
conditions.}
\centering %
\begin{tabular}{rccc}
\hline 
 & Center  & Up  & Down \tabularnewline
\hline 
Center  & 1/4  & 1/8  & 1/8 \tabularnewline
Up  & 1/8  & 1/16  & 1/16 \tabularnewline
Down  & 1/8  & 1/16  & 1/16 \tabularnewline
\hline 
\end{tabular}\label{tab:allocation} 
\end{table}

For each session, each trial was randomly assigned to one of the conditions
in accordance with Table~\ref{tab:allocation}. The number of experimental
sessions was chosen so that the expected number of (non-practice)
trials in the conditions with lowest probabilities was at least $300$.
This number of observations was chosen based on Refs.~\cite{Cepeda-Cuervo.2008.Intervalos},
whose results show that coverage errors with respect to nominal values
are below $1\%$ for almost all confidence intervals for proportions
with $n>300$.

\section{Analyses}

\label{sec:analyses}

Based on the experimental design depicted in Table \ref{tab:allocation},
we specify the following properties:
\begin{itemize}
\item {\footnotesize{}$l_{c}$: a dot is presented in the center of the
left circle;}{\footnotesize \par}
\item {\footnotesize{}$r_{c}$: a dot is presented in the center of the
right circle;}{\footnotesize \par}
\item {\footnotesize{}$l_{u}$: a dot is presented above the center of the
left circle;}{\footnotesize \par}
\item {\footnotesize{}$r_{u}$: a dot is presented above the center of the
right circle;}{\footnotesize \par}
\item {\footnotesize{}$l_{d}$: a dot is presented below the center of the
left circle; and}{\footnotesize \par}
\item {\footnotesize{}$r_{d}$: a dot is presented below the center of the
right circle.}{\footnotesize \par}
\end{itemize}
The 9 experimental conditions (contexts) then are denoted $l_{c}r_{c},l_{c}r_{u}$,
etc. Thus, the system of measurements depicted by the matrix in Figure
\ref{fig:positionSystem} represents the complete $3\times3$ design
given in Table \ref{tab:allocation}.

\begin{figure}[H]
\[
\boxed{\begin{array}{lcccccc}
 & l_{c} & r_{c} & l_{u} & r_{u} & l_{d} & r_{d}\\
l_{c}r_{c} & R_{l_{c}}^{l_{c}r_{c}} & R_{r_{c}}^{l_{c}r_{c}} & \cdot & \cdot & \cdot & \cdot\\
l_{u}r_{c} & \cdot & R_{r_{c}}^{l_{u}r_{c}} & R_{l_{u}}^{l_{u}r_{c}} & \cdot & \cdot & \cdot\\
l_{u}r_{u} & \cdot & \cdot & R_{l_{u}}^{l_{u}r_{u}} & R_{r_{u}}^{l_{u}r_{u}} & \cdot & \cdot\\
l_{d}r_{u} & \cdot & \cdot & \cdot & R_{r_{u}}^{l_{d}r_{u}} & R_{l_{d}}^{l_{d}r_{u}} & \cdot\\
l_{d}r_{d} & \cdot & \cdot & \cdot & \cdot & R_{l_{d}}^{l_{d}r_{d}} & R_{r_{d}}^{l_{d}r_{d}}\\
l_{c}r_{u} & R_{l_{c}}^{l_{c}r_{u}} & \cdot & \cdot & R_{r_{u}}^{l_{c}r_{u}} & \cdot & \cdot\\
l_{u}r_{d} & \cdot & \cdot & R_{l_{u}}^{l_{u}r_{d}} & \cdot & \cdot & R_{r_{d}}^{l_{u}r_{d}}\\
l_{d}r_{c} & \cdot & R_{r_{c}}^{l_{d}r_{c}} & \cdot & \cdot & R_{l_{d}}^{l_{d}r_{c}} & \cdot\\
l_{c}r_{d} & R_{l_{c}}^{l_{c}r_{d}} & \cdot & \cdot & \cdot & \cdot & R_{r_{d}}^{l_{c}r_{d}}
\end{array}}
\]

\caption{System of measurements for double detection experiment.\label{fig:positionSystem}}
\end{figure}

We approach the exploration of this system through the theory of contextuality
for cyclic systems in two ways. Firstly, note that from the system
in Figure \ref{fig:positionSystem} we can extract six different cyclic
subsystems of rank 6 and nine of rank 4. One of the rank 4 subsystems
is presented in the left matrix in Figure \ref{fig:examplesSubsystems}.
One of the rank 6 subsystems is shown in the right matrix in Figure
\ref{fig:examplesSubsystems}.

\begin{figure}[H]
\[
\boxed{\begin{array}{lcccc}
 & l_{c} & r_{c} & l_{u} & r_{u}\\
l_{c}r_{c} & R_{l_{c}}^{l_{c}r_{c}} & R_{r_{c}}^{l_{c}r_{c}} & \cdot & \cdot\\
l_{u}r_{c} & \cdot & R_{r_{c}}^{l_{u}r_{c}} & R_{l_{u}}^{l_{u}r_{c}} & \cdot\\
l_{u}r_{u} & \cdot & \cdot & R_{l_{u}}^{l_{u}r_{u}} & R_{r_{u}}^{l_{u}r_{u}}\\
l_{c}r_{u} & R_{l_{c}}^{l_{c}r_{u}} & \cdot & \cdot & R_{r_{u}}^{l_{c}r_{u}}
\end{array}}\qquad\boxed{\begin{array}{lcccccc}
 & l_{c} & r_{c} & l_{u} & r_{u} & l_{d} & r_{d}\\
l_{c}r_{c} & R_{l_{c}}^{l_{c}r_{c}} & R_{r_{c}}^{l_{c}r_{c}} & \cdot & \cdot & \cdot & \cdot\\
l_{u}r_{c} & \cdot & R_{r_{c}}^{l_{u}r_{c}} & R_{l_{u}}^{l_{u}r_{c}} & \cdot & \cdot & \cdot\\
l_{u}r_{u} & \cdot & \cdot & R_{l_{u}}^{l_{u}r_{u}} & R_{r_{u}}^{l_{u}r_{u}} & \cdot & \cdot\\
l_{d}r_{u} & \cdot & \cdot & \cdot & R_{r_{u}}^{l_{d}r_{u}} & R_{l_{d}}^{l_{d}r_{u}} & \cdot\\
l_{d}r_{d} & \cdot & \cdot & \cdot & \cdot & R_{l_{d}}^{l_{d}r_{d}} & R_{r_{d}}^{l_{d}r_{d}}\\
l_{c}r_{d} & R_{l_{c}}^{l_{c}r_{d}} & \cdot & \cdot & \cdot & \cdot & R_{r_{d}}^{l_{c}r_{d}}
\end{array}}
\]
\caption{Examples of cyclic subsystems of rank 4 and 6.\label{fig:examplesSubsystems}}
\end{figure}

Secondly, in addition to the definition of the quantities as presented
above, there are several interesting systems produced by redefining
these quantities.\footnote{There are also several uninteresting ways to construct systems of
measurements for the conditions and measurements in this experiment.
Examples of how to construct them and why they are not interesting
may be found in Ref. \cite{Dzhafarov.2015.Context-content}} From the description of the double-detection paradigm, one can argue,
e.g., that the center location may be viewed as a signal to be detected,
with either of the two off-center locations being treated as absence
of the signal. This way of looking at the stimuli induces the following
definition of the properties to be measured:
\begin{itemize}
\item {\footnotesize{}$l_{c}$: a dot is presented in the center of the
left circle;}{\footnotesize \par}
\item {\footnotesize{}$r_{c}$: a dot is presented in the center of the
right circle;}{\footnotesize \par}
\item {\footnotesize{}$l_{ud}$: a dot is presented off-center in the left
circle;}{\footnotesize \par}
\item {\footnotesize{}$r_{ud}$: a dot is presented off-center in the right
circle.}{\footnotesize \par}
\end{itemize}
Analogously one could also consider {\footnotesize{}$l_{cu}$}, {\footnotesize{}$l_{cd}$},
{\footnotesize{}$r_{cu}$}, {\footnotesize{}$r_{cd}$}, as properties
to be measured in appropriately chosen contexts,

Another way of dealing with our data is to consider the locations
of the dots as properties to be measured (by responses attributing
to them to a left or to a right circle). For instance, a pair of properties
can be chosen as 
\begin{itemize}
\item {\footnotesize{}$c$: a dot is presented in the center of a circle;
and}{\footnotesize \par}
\item {\footnotesize{}$ud$: a dot is presented off the center of a circle.}{\footnotesize \par}
\end{itemize}
A systematic application of both of these redefinitions leads to also
consider quantities {\footnotesize{}$l_{cu}$}, {\footnotesize{}$l_{cd}$},
{\footnotesize{}$r_{cu}$}, {\footnotesize{}$r_{cd}$}, {\footnotesize{}$u$},
{\footnotesize{}$cd$}, {\footnotesize{}$d$}, and {\footnotesize{}$cu$}
with the analogous interpretations. In this way, six systems of rank
2 and 27 systems of rank 4 may be constructed. Thus, we shall consider
systems with the structures depicted by the matrices in Figures \ref{fig:rank2},
\ref{fig:rank4}, and \ref{fig:rank6}.

\begin{figure}[tbph]
\[
\boxed{\begin{array}{lcc}
 & x & y\\
l_{x}r_{y} & R_{x}^{l_{x}r_{y}} & R_{y}^{l_{x}r_{y}}\\
l_{y}r_{x} & R_{x}^{l_{y}r_{x}} & R_{y}^{l_{y}r_{x}}
\end{array}}
\]
\caption{Rank 2 systems structure where $\left(x,y\right)$ is any of $\left(c,ud\right),\left(cu,d\right),\left(cd,u\right),\left(c,u\right),\left(c,d\right),\left(u,d\right).$\label{fig:rank2}}
\end{figure}

\begin{figure}[tbph]
\[
\boxed{\begin{array}{lcccc}
 & l_{x} & r_{x} & l_{y} & r_{y}\\
l_{x}r_{x} & R_{l_{x}}^{l_{x}r_{x}} & R_{r_{x}}^{l_{x}r_{x}} & \cdot & \cdot\\
l_{y}r_{x} & \cdot & R_{r_{x}}^{l_{y}r_{x}} & R_{l_{y}}^{l_{y}r_{x}} & \cdot\\
l_{y}r_{y} & \cdot & \cdot & R_{l_{y}}^{l_{y}r_{y}} & R_{r_{y}}^{l_{y}r_{y}}\\
l_{x}r_{y} & R_{l_{x}}^{l_{x}r_{y}} & \cdot & \cdot & R_{r_{y}}^{l_{x}r_{y}}
\end{array}}
\]
\caption{Rank 4 systems structure where $\left(l_{x},l_{y}\right)$ is any
of $\left(l_{c},l_{ud}\right),\left(l_{cu},l_{d}\right),\left(l_{cd},l_{u}\right),\left(l_{c},l_{u}\right),\left(l_{c},l_{d}\right),\left(l_{u},l_{d}\right)$,
and $\left(r_{x},r{}_{y}\right)$ is any of $\left\{ \left(r_{c},r{}_{ud}\right),\left(r_{cu},r_{d}\right),\left(r_{cd},r_{u}\right),\left(r_{c},r_{u}\right),\left(r_{c},r_{d}\right),\left(r_{u},r_{d}\right)\right\} .$\label{fig:rank4}}
\end{figure}

\begin{figure}[tbph]
\[
\boxed{\begin{array}{lcccccc}
 & l_{x} & r_{x} & l_{y} & r_{y} & l_{z} & r_{z}\\
l_{x}r_{x} & R_{l_{x}}^{l_{x}r_{x}} & R_{r_{x}}^{l_{x}r_{x}} & \cdot & \cdot & \cdot & \cdot\\
l_{y}r_{x} & \cdot & R_{r_{x}}^{l_{y}r_{x}} & R_{l_{y}}^{l_{y}r_{x}} & \cdot & \cdot & \cdot\\
l_{y}r_{y} & \cdot & \cdot & R_{l_{y}}^{l_{y}r_{y}} & R_{r_{y}}^{l_{y}r_{y}} & \cdot & \cdot\\
l_{z}r_{y} & \cdot & \cdot & \cdot & R_{r_{y}}^{l_{z}r_{y}} & R_{l_{z}}^{l_{z}r_{y}} & \cdot\\
l_{z}r_{z} & \cdot & \cdot & \cdot & \cdot & R_{l_{z}}^{l_{z}r_{z}} & R_{r_{z}}^{l_{z}r_{z}}\\
l_{x}r_{z} & R_{l_{x}}^{l_{x}r_{z}} & \cdot & \cdot & \cdot & \cdot & R_{r_{z}}^{l_{x}r_{z}}
\end{array}}
\]
\caption{Rank 6 systems structure where $\left(x,y,z\right)$ is any of $\left(c,u,d\right),\left(c,d,u\right),\left(u,c,d\right),\left(d,c,u\right),\left(u,d,c\right),\left(d,u,c\right).$\label{fig:rank6}}
\end{figure}

\section{Results}

\label{sec:results}

\subsection{Results for Cyclic Subsystems}

Table~\ref{tab:subjectData} presents the individual data for all
of the expectations used in the calculations of all subsystems. Note
that the statistics associated with the redefined quantities are obtained
by an apropriate linear combination of those in Table~\ref{tab:subjectData}
with weights proportional to the number of trials of the combined
conditions.

\begin{table}[tbph]
\caption{Individual level data}
\centering %
\begin{tabular}{|cc|rrr|rrr|rrr|}
\hline 
\multicolumn{2}{|c|}{} & \multicolumn{3}{c|}{$P1$} & \multicolumn{3}{c|}{$P2$} & \multicolumn{3}{c|}{$P3$}\tabularnewline
\hline 
$l$  & $r$  & $\left\langle R_{l}^{lr}\right\rangle $  & $\left\langle R_{r}^{lr}\right\rangle $  & $\left\langle R_{l}^{lr}R_{r}^{lr}\right\rangle $  & $\left\langle R_{l}^{lr}\right\rangle $  & $\left\langle R_{r}^{lr}\right\rangle $  & $\left\langle R_{l}^{lr}R_{r}^{lr}\right\rangle $  & $\left\langle R_{l}^{lr}\right\rangle $  & $\left\langle R_{r}^{lr}\right\rangle $  & $\left\langle R_{l}^{lr}R_{r}^{lr}\right\rangle $ \tabularnewline
\hline 
$l_{c}$  & $r_{c}$  & $0.4349$  & $0.2730$  & $0.4825$  & $0.7317$  & $0.5683$  & $0.3984$  & $0.3582$  & $0.1946$  & $-0.0913$ \tabularnewline
$l_{c}$  & $r_{u}$  & $0.6190$  & $-0.5397$  & $-0.2095$  & $0.7016$  & $-0.0825$  & $-0.2413$  & $0.6762$  & $-0.8508$  & $-0.6159$ \tabularnewline
$l_{c}$  & $r_{d}$  & $-0.1873$  & $0.2698$  & $0.4095$  & $0.8857$  & $-0.8635$  & $-0.7937$  & $0.3937$  & $-0.3524$  & $-0.3429$ \tabularnewline
$l_{u}$  & $r_{c}$  & $-0.5048$  & $0.1175$  & $0.2254$  & $-0.2063$  & $0.5238$  & $-0.5302$  & $-0.7302$  & $0.6603$  & $-0.5683$ \tabularnewline
$l_{u}$  & $r_{u}$  & $0.0476$  & $-0.0286$  & $0.4794$  & $0.1111$  & $0.1683$  & $0.2190$  & $-0.4904$  & $-0.6624$  & $0.4459$ \tabularnewline
$l_{u}$  & $r_{d}$  & $-0.8476$  & $-0.0857$  & $0.1619$  & $0.2254$  & $-0.7778$  & $-0.4222$  & $-0.7643$  & $-0.2166$  & $0.0446$ \tabularnewline
$l_{d}$  & $r_{c}$  & $0.5873$  & $-0.3937$  & $-0.0825$  & $-0.6667$  & $0.7810$  & $-0.5238$  & $-0.4159$  & $0.3429$  & $-0.4762$ \tabularnewline
$l_{d}$  & $r_{u}$  & $0.5619$  & $-0.9111$  & $-0.5365$  & $-0.7333$  & $0.2635$  & $-0.4286$  & $-0.2508$  & $-0.7079$  & $0.0095$ \tabularnewline
$l_{d}$  & $r_{d}$  & $0.5111$  & $0.3016$  & $0.4730$  & $-0.5175$  & $-0.5937$  & $0.5810$  & $-0.3079$  & $-0.1746$  & $0.0413$ \tabularnewline
\hline 
\end{tabular}\label{tab:subjectData} 
\end{table}

Table~\ref{tab:rank6cyclic} presents the values of $\Lambda{C}$,
\ICC{}, and $\Delta{C}$ calculated for each participant and each
of the rank 6 cyclic subsystems. Table~\ref{tab:rank4cyclic} presents
the respective values for each of the rank 4 cyclic subsystems. For
all participants, the subsystems are noncontextual.

\begin{table}[tbph]
\centering{}\caption{Contextuality cyclic subsystems of rank 6}
{\scriptsize{}}%
\begin{tabular}{|c|ccc|ccc|ccc|}
\hline 
{\scriptsize{}System } & \multicolumn{3}{c|}{{\scriptsize{}$P1$}} & \multicolumn{3}{c|}{{\scriptsize{}$P2$}} & \multicolumn{3}{c|}{{\scriptsize{}$P3$}}\tabularnewline
\cline{2-10} 
{\scriptsize{}$(l_{x,},l_{y},l_{z}),(r{}_{x},r_{y},r_{z})$ } & {\scriptsize{}$\Lambda{C}$ } & {\scriptsize{}ICC } & {\scriptsize{}$\Delta{C}$ } & {\scriptsize{}$\Lambda{C}$ } & {\scriptsize{}ICC} & {\scriptsize{}$\Delta{C}$ } & {\scriptsize{}$\Lambda{C}$ } & {\scriptsize{}ICC} & {\scriptsize{}$\Delta{C}$ }\tabularnewline
\hline 
{\scriptsize{}$(l_{c},l_{d},l_{u}),(r{}_{d},r_{u},r_{c})$} & {\scriptsize{}1.6254 } & {\scriptsize{}2.4127 } & {\scriptsize{}-4.7873 } & {\scriptsize{}2.4571 } & {\scriptsize{}1.3714 } & {\scriptsize{}-2.9143 } & {\scriptsize{}2.0382 } & {\scriptsize{}1.0779 } & {\scriptsize{}-3.0397 }\tabularnewline
{\scriptsize{}$(l_{d},l_{c},l_{u}),(r{}_{c},r_{d},r_{u})$} & {\scriptsize{}1.7143 } & {\scriptsize{}2.4889 } & {\scriptsize{}-4.7746 } & {\scriptsize{}2.4508 } & {\scriptsize{}1.4286 } & {\scriptsize{}-2.9778 } & {\scriptsize{}2.4078 } & {\scriptsize{}1.3138 } & {\scriptsize{}-2.9060 }\tabularnewline
{\scriptsize{}$(l_{d},l_{u},l_{c}),(r{}_{c},r_{u},r_{d})$} & {\scriptsize{}1.9873 } & {\scriptsize{}3.4476 } & {\scriptsize{}-5.4603 } & {\scriptsize{}2.3476 } & {\scriptsize{}0.7286 } & {\scriptsize{}-2.3810 } & {\scriptsize{}1.4104 } & {\scriptsize{}0.8040 } & {\scriptsize{}-3.3936 }\tabularnewline
{\scriptsize{}$(l_{c},l_{u},l_{d}),(r{}_{d},r_{c},r_{u})$} & {\scriptsize{}2.6063 } & {\scriptsize{}2.2952 } & {\scriptsize{}-3.6889 } & {\scriptsize{}2.9508 } & {\scriptsize{}1.0968 } & {\scriptsize{}-2.1460 } & {\scriptsize{}1.4991 } & {\scriptsize{}1.0213 } & {\scriptsize{}-3.5222 }\tabularnewline
{\scriptsize{}$(l_{u},l_{c},l_{d}),(r{}_{d},r_{u},r_{c})$} & {\scriptsize{}1.7238 } & {\scriptsize{}2.7206 } & {\scriptsize{}-4.9968 } & {\scriptsize{}2.3857 } & {\scriptsize{}0.9413 } & {\scriptsize{}-2.5556 } & {\scriptsize{}1.7151 } & {\scriptsize{}1.0784 } & {\scriptsize{}-3.3633 }\tabularnewline
{\scriptsize{}$(l_{u},l_{d},l_{c}),(r{}_{c},r_{d},r_{u})$} & {\scriptsize{}1.7651 } & {\scriptsize{}1.4921 } & {\scriptsize{}-3.7270 } & {\scriptsize{}2.1190 } & {\scriptsize{}1.2524 } & {\scriptsize{}-3.1333 } & {\scriptsize{}1.3708 } & {\scriptsize{}1.0598 } & {\scriptsize{}-3.6890 }\tabularnewline
\hline 
\end{tabular}\label{tab:rank6cyclic} 
\end{table}

\begin{table}[tbph]
\centering{}\caption{Contextuality cyclic subsystems of rank 4}
{\scriptsize{}}%
\begin{tabular}{|c|ccc|ccc|ccc|}
\hline 
{\scriptsize{}System } & \multicolumn{3}{c|}{{\scriptsize{}$P1$}} & \multicolumn{3}{c|}{{\scriptsize{}$P2$}} & \multicolumn{3}{c|}{{\scriptsize{}$P3$}}\tabularnewline
\cline{2-10} 
{\scriptsize{}$(l_{x},l_{y}),(r{}_{x},r_{y})$ } & {\scriptsize{}$\Lambda{C}$ } & {\scriptsize{}ICC } & {\scriptsize{}$\Delta{C}$ } & {\scriptsize{}$\Lambda{C}$ } & {\scriptsize{}ICC} & {\scriptsize{}$\Delta{C}$ } & {\scriptsize{}$\Lambda{C}$ } & {\scriptsize{}ICC} & {\scriptsize{}$\Delta{C}$ }\tabularnewline
\hline 
{\scriptsize{}$(l_{c},l_{u}),(r{}_{c},r_{u})$} & {\scriptsize{}1.3968 } & {\scriptsize{}1.4032 } & {\scriptsize{}-2.0063 } & {\scriptsize{}0.9508 } & {\scriptsize{}0.6429 } & {\scriptsize{}-1.6921 } & {\scriptsize{}1.7213 } & {\scriptsize{}1.2118 } & {\scriptsize{}-1.4904 }\tabularnewline
{\scriptsize{}$(l_{c},l_{u}),(r{}_{c},r_{d})$} & {\scriptsize{}0.9556 } & {\scriptsize{}1.4762 } & {\scriptsize{}-2.5206 } & {\scriptsize{}2.1444 } & {\scriptsize{}0.7159 } & {\scriptsize{}-0.5714 } & {\scriptsize{}1.0470 } & {\scriptsize{}0.6711 } & {\scriptsize{}-1.6241 }\tabularnewline
{\scriptsize{}$(l_{c},l_{u}),(r{}_{u},r_{d})$} & {\scriptsize{}1.2603 } & {\scriptsize{}2.5683 } & {\scriptsize{}-3.3079 } & {\scriptsize{}1.6762 } & {\scriptsize{}0.6349 } & {\scriptsize{}-0.9587 } & {\scriptsize{}1.3600 } & {\scriptsize{}0.8806 } & {\scriptsize{}-1.5206 }\tabularnewline
{\scriptsize{}$(l_{c},l_{d}),(r{}_{c},r_{u})$} & {\scriptsize{}1.3111 } & {\scriptsize{}1.2476 } & {\scriptsize{}-1.9365 } & {\scriptsize{}1.5921 } & {\scriptsize{}0.6556 } & {\scriptsize{}-1.0635 } & {\scriptsize{}1.1929 } & {\scriptsize{}0.7742 } & {\scriptsize{}-1.5812 }\tabularnewline
{\scriptsize{}$(l_{c},l_{d}),(r{}_{c},r_{d})$} & {\scriptsize{}1.4476 } & {\scriptsize{}1.3968 } & {\scriptsize{}-1.9492 } & {\scriptsize{}1.5000 } & {\scriptsize{}0.7857 } & {\scriptsize{}-1.2857 } & {\scriptsize{}0.9517 } & {\scriptsize{}0.4694 } & {\scriptsize{}-1.5177 }\tabularnewline
{\scriptsize{}$(l_{c},l_{d}),(r{}_{u},r_{d})$} & {\scriptsize{}1.2095 } & {\scriptsize{}1.2603 } & {\scriptsize{}-2.0508 } & {\scriptsize{}2.0444 } & {\scriptsize{}1.0159 } & {\scriptsize{}-0.9714 } & {\scriptsize{}0.9905 } & {\scriptsize{}0.6603 } & {\scriptsize{}-1.6698 }\tabularnewline
{\scriptsize{}$(l_{u},l_{d}),(r{}_{c},r_{u})$} & {\scriptsize{}1.1587 } & {\scriptsize{}1.9714 } & {\scriptsize{}-2.8127 } & {\scriptsize{}1.7016 } & {\scriptsize{}0.7365 } & {\scriptsize{}-1.0349 } & {\scriptsize{}1.4808 } & {\scriptsize{}0.7678 } & {\scriptsize{}-1.2870 }\tabularnewline
{\scriptsize{}$(l_{u},l_{d}),(r{}_{c},r_{d})$} & {\scriptsize{}0.9429 } & {\scriptsize{}1.3175 } & {\scriptsize{}-2.3746 } & {\scriptsize{}2.0571 } & {\scriptsize{}1.0222 } & {\scriptsize{}-0.9651 } & {\scriptsize{}1.0478 } & {\scriptsize{}0.5015 } & {\scriptsize{}-1.4538 }\tabularnewline
{\scriptsize{}$(l_{u},l_{d}),(r{}_{u},r_{d})$} & {\scriptsize{}1.6508 } & {\scriptsize{}2.2159 } & {\scriptsize{}-2.5651 } & {\scriptsize{}1.2127 } & {\scriptsize{}0.6095 } & {\scriptsize{}-1.3968 } & {\scriptsize{}0.5222 } & {\scriptsize{}0.4185 } & {\scriptsize{}-1.8963 }\tabularnewline
\hline 
\end{tabular}\label{tab:rank4cyclic} 
\end{table}

\subsection{Results for Cyclic Systems with Redefined Quantities}

Table~\ref{tab:rank2redefined} presents the values of $\Lambda{C}$,
\ICC{}, and $\Delta{C}$ calculated for each participant for each
of the rank 2 cyclic systems, and Table~\ref{tab:rank4redefined}
shows those for the rank 4 cyclic systems. Note that for participant
$P3$, two of the rank 2 systems, those with $\left(x,y\right)=\left(c,d\right)$
and $\left(x,y\right)=\left(cu,d\right)$, have a positive $\Delta{C}$
value, which might suggest that these two systems show contextuality.
However, their respective confidence intervals, $\Delta{C}_{(cu,d)}\in(-0.267,0.241)$
and $\Delta{C}_{(c,d)}\in(-0.233,0.215)$,\footnote{$95\%$ confidence intervals corrected by Bonferroni for the number
of tests for $\Delta{C}$ values in the experiment. However, it should
be noted that even uncorrected intervals covered the value $0$.} indicate that the values are consistent with lack of contextuality.

{\scriptsize{}}
\begin{table}[tbph]

\centering{}{\scriptsize{}\caption{Contextuality cyclic systems of rank 2\label{tab:rank2redefined}}
}%
\begin{tabular}{|c|ccc|ccc|ccc|}
\hline 
{\scriptsize{}System } & \multicolumn{3}{c|}{{\scriptsize{}$P1$}} & \multicolumn{3}{c|}{{\scriptsize{}$P2$}} & \multicolumn{3}{c|}{{\scriptsize{}$P3$}}\tabularnewline
\cline{2-10} 
{\scriptsize{}$(x_{,}y)$ } & {\scriptsize{}$\Lambda{C}$ } & {\scriptsize{}ICC } & {\scriptsize{}$\Delta{C}$ } & {\scriptsize{}$\Lambda{C}$ } & {\scriptsize{}ICC} & {\scriptsize{}$\Delta{C}$ } & {\scriptsize{}$\Lambda{C}$ } & {\scriptsize{}ICC} & {\scriptsize{}$\Delta{C}$ }\tabularnewline
\hline 
{\scriptsize{}$(c,ud)$ } & {\scriptsize{}0.0286 } & {\scriptsize{}0.5302 } & {\scriptsize{}-0.5016 } & {\scriptsize{}0.0095 } & {\scriptsize{}0.1778 } & {\scriptsize{}-0.1683 } & {\scriptsize{}0.0429 } & {\scriptsize{}0.0619 } & {\scriptsize{}-0.0190 }\tabularnewline
{\scriptsize{}$(cd,u)$ } & {\scriptsize{}0.5228 } & {\scriptsize{}0.5947 } & {\scriptsize{}-0.0720 } & {\scriptsize{}0.1905 } & {\scriptsize{}0.2286 } & {\scriptsize{}-0.0381 } & {\scriptsize{}0.0430 } & {\scriptsize{}0.0631 } & {\scriptsize{}-0.0201 }\tabularnewline
{\scriptsize{}$(cu,d)$ } & {\scriptsize{}0.5608 } & {\scriptsize{}0.5862 } & {\scriptsize{}-0.0254 } & {\scriptsize{}0.1778 } & {\scriptsize{}0.2032 } & {\scriptsize{}-0.0254 } & {\scriptsize{}0.1003 } & {\scriptsize{}0.0695 } & {\scriptsize{}0.0308 }\tabularnewline
{\scriptsize{}$(c,u)$ } & {\scriptsize{}0.4349 } & {\scriptsize{}0.5365 } & {\scriptsize{}-0.1016 } & {\scriptsize{}0.2889 } & {\scriptsize{}0.3016 } & {\scriptsize{}-0.0127 } & {\scriptsize{}0.0476 } & {\scriptsize{}0.1365 } & {\scriptsize{}-0.0889 }\tabularnewline
{\scriptsize{}$(c,d)$ } & {\scriptsize{}0.4921 } & {\scriptsize{}0.5238 } & {\scriptsize{}-0.0317 } & {\scriptsize{}0.2698 } & {\scriptsize{}0.3016 } & {\scriptsize{}-0.0317 } & {\scriptsize{}0.1333 } & {\scriptsize{}0.1143 } & {\scriptsize{}0.0190 }\tabularnewline
{\scriptsize{}$(u,d)$ } & {\scriptsize{}0.6984 } & {\scriptsize{}0.7111 } & {\scriptsize{}-0.0127 } & {\scriptsize{}0.0063 } & {\scriptsize{}0.0825 } & {\scriptsize{}-0.0762 } & {\scriptsize{}0.0351 } & {\scriptsize{}0.0906 } & {\scriptsize{}-0.0556 }\tabularnewline
\hline 
\end{tabular}{\scriptsize \par}
\end{table}
{\scriptsize \par}

\begin{table}[tbph]
\centering{}\caption{Contextuality cyclic systems of rank 4}
{\scriptsize{}}%
\begin{tabular}{|c|ccc|ccc|ccc|}
\hline 
{\scriptsize{}System } & \multicolumn{3}{c|}{{\scriptsize{}$P1$}} & \multicolumn{3}{c|}{{\scriptsize{}$P2$}} & \multicolumn{3}{c|}{{\scriptsize{}$P3$}}\tabularnewline
\cline{2-10} 
{\scriptsize{}$(l_{x},l_{y}),(r{}_{x},r_{y})$ } & {\scriptsize{}$\Lambda{C}$ } & {\scriptsize{}ICC } & {\scriptsize{}$\Delta{C}$ } & {\scriptsize{}$\Lambda{C}$ } & {\scriptsize{}ICC} & {\scriptsize{}$\Delta{C}$ } & {\scriptsize{}$\Lambda{C}$ } & {\scriptsize{}ICC} & {\scriptsize{}$\Delta{C}$ }\tabularnewline
\hline 
{\scriptsize{}$(l_{c},l_{ud}),(r{}_{c},r_{ud})$} & {\scriptsize{}0.6556 } & {\scriptsize{}0.7032 } & {\scriptsize{}-2.0476 } & {\scriptsize{}1.4556 } & {\scriptsize{}0.5921 } & {\scriptsize{}-1.1365 } & {\scriptsize{}1.2281 } & {\scriptsize{}0.7648 } & {\scriptsize{}-1.5367 }\tabularnewline
{\scriptsize{}$(l_{c},l_{ud}),(r{}_{cd},r_{u})$} & {\scriptsize{}0.7926 } & {\scriptsize{}1.1228 } & {\scriptsize{}-2.3302 } & {\scriptsize{}0.6720 } & {\scriptsize{}0.5238 } & {\scriptsize{}-1.8519 } & {\scriptsize{}1.3525 } & {\scriptsize{}0.9192 } & {\scriptsize{}-1.5667 }\tabularnewline
{\scriptsize{}$(l_{c},l_{ud}),(r{}_{cu},r_{d})$} & {\scriptsize{}0.9407 } & {\scriptsize{}1.3937 } & {\scriptsize{}-2.4529 } & {\scriptsize{}1.2857 } & {\scriptsize{}0.7460 } & {\scriptsize{}-1.4603 } & {\scriptsize{}0.9247 } & {\scriptsize{}0.5181 } & {\scriptsize{}-1.5934 }\tabularnewline
{\scriptsize{}$(l_{c},l_{ud}),(r{}_{c},r_{u})$} & {\scriptsize{}0.7349 } & {\scriptsize{}0.9286 } & {\scriptsize{}-2.1937 } & {\scriptsize{}1.2714 } & {\scriptsize{}0.5381 } & {\scriptsize{}-1.2667 } & {\scriptsize{}1.4568 } & {\scriptsize{}0.9931 } & {\scriptsize{}-1.5363 }\tabularnewline
{\scriptsize{}$(l_{c},l_{ud}),(r{}_{c},r_{d})$} & {\scriptsize{}1.1381 } & {\scriptsize{}1.4048 } & {\scriptsize{}-2.2667 } & {\scriptsize{}1.6397 } & {\scriptsize{}0.7063 } & {\scriptsize{}-1.0667 } & {\scriptsize{}0.9993 } & {\scriptsize{}0.5365 } & {\scriptsize{}-1.5371 }\tabularnewline
{\scriptsize{}$(l_{c},l_{ud}),(r{}_{u},r_{d})$} & {\scriptsize{}0.9079 } & {\scriptsize{}1.5111 } & {\scriptsize{}-2.6032 } & {\scriptsize{}1.2190 } & {\scriptsize{}0.8254 } & {\scriptsize{}-1.6063 } & {\scriptsize{}1.1431 } & {\scriptsize{}0.7703 } & {\scriptsize{}-1.6271 }\tabularnewline
{\scriptsize{}$(l_{cd},l_{u}),(r{}_{c},r_{ud})$} & {\scriptsize{}0.7841 } & {\scriptsize{}0.4688 } & {\scriptsize{}-1.6847 } & {\scriptsize{}1.0423 } & {\scriptsize{}0.6106 } & {\scriptsize{}-1.5683 } & {\scriptsize{}1.3443 } & {\scriptsize{}0.7911 } & {\scriptsize{}-1.4469 }\tabularnewline
{\scriptsize{}$(l_{cd},l_{u}),(r{}_{cd},r_{u})$} & {\scriptsize{}1.3418 } & {\scriptsize{}1.6402 } & {\scriptsize{}-2.2984 } & {\scriptsize{}1.0681 } & {\scriptsize{}0.4804 } & {\scriptsize{}-1.4123 } & {\scriptsize{}1.4357 } & {\scriptsize{}0.9428 } & {\scriptsize{}-1.5070 }\tabularnewline
{\scriptsize{}$(l_{cd},l_{u}),(r{}_{cu},r_{d})$} & {\scriptsize{}0.8127 } & {\scriptsize{}1.6275 } & {\scriptsize{}-2.8148 } & {\scriptsize{}0.9975 } & {\scriptsize{}0.5284 } & {\scriptsize{}-1.5309 } & {\scriptsize{}0.7726 } & {\scriptsize{}0.5453 } & {\scriptsize{}-1.7727 }\tabularnewline
{\scriptsize{}$(l_{cd},l_{u}),(r{}_{c},r_{u})$} & {\scriptsize{}1.3175 } & {\scriptsize{}1.3683 } & {\scriptsize{}-2.0508 } & {\scriptsize{}0.9619 } & {\scriptsize{}0.6106 } & {\scriptsize{}-1.6487 } & {\scriptsize{}1.6412 } & {\scriptsize{}1.0639 } & {\scriptsize{}-1.4227 }\tabularnewline
{\scriptsize{}$(l_{cd},l_{u}),(r{}_{c},r_{d})$} & {\scriptsize{}0.7884 } & {\scriptsize{}1.2159 } & {\scriptsize{}-2.4275 } & {\scriptsize{}1.3788 } & {\scriptsize{}0.7037 } & {\scriptsize{}-1.3249 } & {\scriptsize{}1.0473 } & {\scriptsize{}0.5867 } & {\scriptsize{}-1.5394 }\tabularnewline
{\scriptsize{}$(l_{cd},l_{u}),(r{}_{u},r_{d})$} & {\scriptsize{}1.3905 } & {\scriptsize{}2.4508 } & {\scriptsize{}-3.0603 } & {\scriptsize{}1.2804 } & {\scriptsize{}0.4487 } & {\scriptsize{}-1.1683 } & {\scriptsize{}1.0235 } & {\scriptsize{}0.6986 } & {\scriptsize{}-1.6751 }\tabularnewline
{\scriptsize{}$(l_{cu},l_{d}),(r{}_{c},r_{ud})$} & {\scriptsize{}0.6212 } & {\scriptsize{}0.9725 } & {\scriptsize{}-2.3513 } & {\scriptsize{}0.9153 } & {\scriptsize{}0.6868 } & {\scriptsize{}-1.7714 } & {\scriptsize{}0.9903 } & {\scriptsize{}0.4030 } & {\scriptsize{}-1.4127 }\tabularnewline
{\scriptsize{}$(l_{cu},l_{d}),(r{}_{cd},r_{u})$} & {\scriptsize{}1.0328 } & {\scriptsize{}1.3848 } & {\scriptsize{}-2.3520 } & {\scriptsize{}0.6603 } & {\scriptsize{}0.6145 } & {\scriptsize{}-1.9541 } & {\scriptsize{}0.8142 } & {\scriptsize{}0.5372 } & {\scriptsize{}-1.7230 }\tabularnewline
{\scriptsize{}$(l_{cu},l_{d}),(r{}_{cu},r_{d})$} & {\scriptsize{}1.3051 } & {\scriptsize{}1.4399 } & {\scriptsize{}-2.1347 } & {\scriptsize{}1.7129 } & {\scriptsize{}0.8698 } & {\scriptsize{}-1.1570 } & {\scriptsize{}0.8240 } & {\scriptsize{}0.2918 } & {\scriptsize{}-1.4677 }\tabularnewline
{\scriptsize{}$(l_{cu},l_{d}),(r{}_{c},r_{u})$} & {\scriptsize{}0.9958 } & {\scriptsize{}1.4889 } & {\scriptsize{}-2.4931 } & {\scriptsize{}1.1291 } & {\scriptsize{}0.6423 } & {\scriptsize{}-1.5132 } & {\scriptsize{}0.9988 } & {\scriptsize{}0.5452 } & {\scriptsize{}-1.5464 }\tabularnewline
{\scriptsize{}$(l_{cu},l_{d}),(r{}_{c},r_{d})$} & {\scriptsize{}1.2794 } & {\scriptsize{}1.3704 } & {\scriptsize{}-2.0910 } & {\scriptsize{}1.6857 } & {\scriptsize{}0.8646 } & {\scriptsize{}-1.1788 } & {\scriptsize{}0.9818 } & {\scriptsize{}0.2608 } & {\scriptsize{}-1.2790 }\tabularnewline
{\scriptsize{}$(l_{cu},l_{d}),(r{}_{u},r_{d})$} & {\scriptsize{}1.3566 } & {\scriptsize{}1.5788 } & {\scriptsize{}-2.2222 } & {\scriptsize{}1.7672 } & {\scriptsize{}0.8804 } & {\scriptsize{}-1.1132 } & {\scriptsize{}0.5084 } & {\scriptsize{}0.5496 } & {\scriptsize{}-2.0412 }\tabularnewline
{\scriptsize{}$(l_{c},l_{u}),(r{}_{c},r_{ud})$} & {\scriptsize{}0.9286 } & {\scriptsize{}0.5571 } & {\scriptsize{}-1.6286 } & {\scriptsize{}1.5476 } & {\scriptsize{}0.6492 } & {\scriptsize{}-1.1016 } & {\scriptsize{}1.3842 } & {\scriptsize{}0.9073 } & {\scriptsize{}-1.5231 }\tabularnewline
{\scriptsize{}$(l_{c},l_{u}),(r{}_{cd},r_{u})$} & {\scriptsize{}1.3513 } & {\scriptsize{}1.7915 } & {\scriptsize{}-2.4402 } & {\scriptsize{}0.9534 } & {\scriptsize{}0.5069 } & {\scriptsize{}-1.5534 } & {\scriptsize{}1.6014 } & {\scriptsize{}1.1021 } & {\scriptsize{}-1.5007 }\tabularnewline
{\scriptsize{}$(l_{c},l_{u}),(r{}_{cu},r_{d})$} & {\scriptsize{}0.8095 } & {\scriptsize{}1.6328 } & {\scriptsize{}-2.8233 } & {\scriptsize{}1.6815 } & {\scriptsize{}0.6296 } & {\scriptsize{}-0.9481 } & {\scriptsize{}0.8847 } & {\scriptsize{}0.6947 } & {\scriptsize{}-1.8101 }\tabularnewline
{\scriptsize{}$(l_{c},l_{d}),(r{}_{c},r_{ud})$ } & {\scriptsize{}0.6333 } & {\scriptsize{}1.1063 } & {\scriptsize{}-2.4730 } & {\scriptsize{}1.3635 } & {\scriptsize{}0.6238 } & {\scriptsize{}-1.2603 } & {\scriptsize{}1.0723 } & {\scriptsize{}0.6218 } & {\scriptsize{}-1.5495 }\tabularnewline
{\scriptsize{}$(l_{c},l_{d}),(r{}_{cd},r_{u})$} & {\scriptsize{}1.1016 } & {\scriptsize{}1.1968 } & {\scriptsize{}-2.0952 } & {\scriptsize{}0.8265 } & {\scriptsize{}0.7757 } & {\scriptsize{}-1.9492 } & {\scriptsize{}1.1043 } & {\scriptsize{}0.7363 } & {\scriptsize{}-1.6320 }\tabularnewline
{\scriptsize{}$(l_{c},l_{d}),(r{}_{cu},r_{d})$} & {\scriptsize{}1.3683 } & {\scriptsize{}1.3513 } & {\scriptsize{}-1.9831 } & {\scriptsize{}1.6815 } & {\scriptsize{}0.8624 } & {\scriptsize{}-1.1810 } & {\scriptsize{}0.9647 } & {\scriptsize{}0.4479 } & {\scriptsize{}-1.4833 }\tabularnewline
{\scriptsize{}$(l_{u},l_{d}),(r{}_{c},r_{ud})$} & {\scriptsize{}0.5968 } & {\scriptsize{}0.9143 } & {\scriptsize{}-2.3175 } & {\scriptsize{}1.2317 } & {\scriptsize{}0.8127 } & {\scriptsize{}-1.5810 } & {\scriptsize{}1.2643 } & {\scriptsize{}0.5585 } & {\scriptsize{}-1.2942 }\tabularnewline
{\scriptsize{}$(l_{u},l_{d}),(r{}_{cd},r_{u})$ } & {\scriptsize{}1.3228 } & {\scriptsize{}1.7608 } & {\scriptsize{}-2.4381 } & {\scriptsize{}1.2974 } & {\scriptsize{}0.6180 } & {\scriptsize{}-1.3206 } & {\scriptsize{}1.1044 } & {\scriptsize{}0.6240 } & {\scriptsize{}-1.5195 }\tabularnewline
{\scriptsize{}$(l_{u},l_{d}),(r{}_{cu},r_{d})$} & {\scriptsize{}1.1788 } & {\scriptsize{}1.6169 } & {\scriptsize{}-2.4381 } & {\scriptsize{}1.7757 } & {\scriptsize{}0.8847 } & {\scriptsize{}-1.1090 } & {\scriptsize{}0.5485 } & {\scriptsize{}0.4365 } & {\scriptsize{}-1.8880 }\tabularnewline
\hline 
\end{tabular}\label{tab:rank4redefined} 
\end{table}

\section{Conclusions}

\label{sec:conclusions}

The experiment presented in this paper illustrates the use of the
double factorial paradigm in the search of contextuality in behavioral
systems, namely in the responses of human observers in a double-detection
task. This paradigm provides the closest analogue in psychophysical
research to the Alice-Bob EPR/Bohm paradigm.

We have found that for the participants in the study there was no
evidence of contextuality in their responses. These results add to
the existing evidence that points towards lack of contextuality in
psychology (cf. Ref. \cite{Dzhafarov.2015.there}.)

\bibliographystyle{splncs03}

\end{document}